# Evidence for nesting driven charge density waves instabilities in a quasi-two-dimensional material: LaAgSb$_2$

BOSAK Alexeï[1], SOULIOU Sofia-Michaela[2,1], FAUGERAS Clément[3], HEID Rolf[2], MOLAS Maciej R.[3,4], CHEN Rong-Yan[5], WANG Nan-Lin[6], POTEMSKI Marek[3,4]
LE TACON Matthieu[2]

[1] European Synchrotron Radiation Facility, 71 av. de Martyrs, Grenoble, 38043, France

[2] Institute for Quantum Materials and Technologies, Karlsruhe Institute of Technology, 76131 Karlsruhe, Germany

[3] Laboratoire National des Champs Magnétiques Intenses, Univ. Grenoble Alpes, CNRS-UPS-INSA-EMFL, 25 av. des Martyrs, 38042 Grenoble, France

[4] Faculty of Physics, University of Warsaw, ul. Pasteura 5, 02-093 Warsaw, Poland

[5] Department of Physics, Beijing Normal University, Beijing 100875, China.

[6] International Center for Quantum Materials, School of Physics, Peking University, Beijing 100871, China

**Since their theoretical prediction by Peierls in the 30s, charge density waves (CDW) have been one of the most commonly encountered electronic phases in low dimensional metallic systems. The instability mechanism originally proposed combines Fermi surface nesting and electron-phonon coupling but is, strictly speaking, only valid in one dimension. In higher dimensions, its relevance is questionable as sharp maxima in the static electronic susceptibility $\chi(q)$ are smeared out, and are, in many cases, unable to account for the periodicity of the observed charge modulations. Here, we investigate the quasi two-dimensional LaAgSb$_2$, which exhibits two CDW transitions, by a combination of diffuse x-ray scattering, inelastic x-ray scattering and *ab initio* calculations. We demonstrate that the CDW formation is driven by phonons softening. The corresponding Kohn anomalies are visualized in 3D through the momentum distribution of the x-ray diffuse scattering intensity. We show that they can be quantitatively accounted for by considering the electronic susceptibility calculated from a Dirac-like band, weighted by anisotropic electron-phonon coupling. This remarkable agreement sheds new light on the importance of Fermi surface nesting in CDW formation.**

Charge density waves (CDW) in solids are electronic ground states arising from an intrinsic instability of a metallic phase against a spatial modulation of the free carrier density, and are generally accompanied with a static distortion of the crystal lattice. They have originally been predicted and reported in low-dimensional metals (*e.g.* blue bronzes [1], organic salts [2] or tri-tellurides [3,4]), but since then they have been reported in a vast variety of compounds encompassing 2D dichalcogenides [5], superconducting cuprates [6,7], or more recently in nickel pnictide superconductors [8]. The existence of CDW was originally proposed by Peierls who showed that a 1D chain is unstable due to a divergence of the static electronic susceptibility $\chi(q)$ at the wave-vector $q = 2k_F$, that perfectly nests two parallel portions of the Fermi surface [9]. Through electron-phonon coupling (EPC), the phonon spectrum softens at $2k_F$, ultimately resulting in a static distortion of the lattice as a mode's energy vanishes. Unperfect nesting in real quasi-1D materials or in higher dimension compounds, on the other hand, rapidly supresses the susceptibility divergence [10], and thereby invalidates the Peierls scenario in the vast majority of CDW materials [11, 12]. Even without resulting in a diverging electronic susceptibility, the presence of partial Fermi-surface nesting in dimension d>1 can enhance locally the EPC and eventually set the stage for the formation of a CDW [13]. Depending on the system, alternative approaches have been proposed to account for the formation of CDWs, which encompass strong momentum [14,15,16] or orbital [17] dependence of the EPC (in combination with strong anharmonic effects [18,19,20,21]) , spin fluctuations [22] or exciton condensation [23].

The family of RAgSb$_2$, where R is a rare earth ion, has attracted growing attention in the last years due to the different low temperature ground states observed in these compounds when varying the R ion [24,25]. Bulk LaAgSb$_2$ is a yet relatively unexplored compound of tetragonal structure (P4/nmm) that hosts two distinct CDW with two critical temperatures of $T_{CDW1}$ = 207 K and $T_{CDW2}$ = 186 K as evidenced by x-ray diffraction, transport, thermal and NMR studies [26, 27]. The two CDWs present at low temperature in LaAgSb$_2$ are aligned along the *a* and *c* axis, with a rather large real space periodicity of ~17 nm (CDW1, τ$_1$ ~ 0.026 a*) and ~6.5 nm (CDW2, τ$_2$ ~ 1/6 c*) and argued to be consistent with Fermi surface nesting [26]. Interestingly, recent magneto-transport [28] and ARPES [29]investigations have indicated that electronic bands in the vicinity of the Fermi energy involved in the formation of CDW1 are of Dirac type, dispersing linearly as a function of momentum as in graphene, albeit with a band velocity twice smaller. The estimated nesting vector ~(0.09±0.04) π/a, is large but relatively close to $\tau_1$. *Ab initio* calculations further suggest that CDW2 is also related to nested parts of the Fermi surface associated with a distinct electronic band [25, 26]. Time-resolved optical measurements revealed two low energy amplitude modes, suggesting that the CDW instability is triggered by the softening of a low-lying acoustic phonon mode [30]. To date, however, the dispersion of the phonons in this system and their possible role in the formation of the two CDW states has not been investigated.

Having in mind the aforementioned considerations regarding Fermi surface nesting driven formation of CDW in dimensions d > 1, we have performed a series of temperature dependent diffuse scattering (DS) and inelastic x-ray scattering (IXS) experiments in order to unveil the CDW formation mechanism in LaAgSb$_2$. The wave vectors of the CDW instability can be directly identified in the normal state DS intensity distribution, and IXS investigations confirm the soft-phonon driven nature of the CDW instabilities. Interestingly, we observe that the complex 3D momentum distribution of the DS intensity accurately follows that of the 'partial' electronic susceptibility arising from the intraband contribution from the linearly dispersing electronic states. These states do not form a complete Dirac cone (as the dispersion is linear only along one k-space direction, and parabolic in the orthogonal direction), but their strong nesting locally enhances the EPC and yields the CDW formation. This provides a textbook example for the CDW formation in higher-dimensional materials and suggests new routes for the possible design of such states through band structure engineering of metallic systems.

Bulk LaAgSb$_2$ crystals were grown by the self-flux method [15]. Laser-cut disks (with *c* axis normal to the disk surface) were etched by nitric acid and immediately after – by concentrated alkaline solution. The resulting samples are lentil-like with thickness ~50 μm and diameter ~200 μm. This process leads to high quality unstrained material, with a size adapted for diffuse scattering or IXS experiments, with negligible surface pollution. Diffraction patterns show only a few powder-like residuals, presumably coming from antimony derivatives.

Reciprocal space mapping was performed using the diffraction station of ID28 ESRF beamline [31], operated at 0.6968 Å wavelength. Frames were collected in shutterless mode with the angular step of 0.05° and maximal available distance to the detector – 414 mm, preliminary data reduction was done by the CrysAlis package [32], final high-quality reconstructions were produced by a locally developed software. The IXS measurements were performed on the main station of ID28 beamline, as described elsewhere [33]. Energy scans were recorded with ~1.5 meV energy resolution and ~0.25 x 0.25 nm$^{-1}$ momentum resolution. In both cases, the sample temperature was controlled by a Cryostream 700 Plus [34].

DFT band structure calculations for the high-temperature tetragonal structure were performed with the mixed-basis pseudopotential method [35, 36] in the local-density approximation [37]. Norm-conserving pseudopotentials were constructed following the description of Vanderbilt [38] and included the semi-core states La-*5s*, La-*5p*, Ag-*4s*, Ag-*4p*, and Sb-*5s* in the valence space. We used experimental lattice parameters taken from [39] ($a$ = 4.3903 Å, $c$ = 10.840 Å), and relaxed the internal atomic positions. In the mixed-basis approach, local functions at atomic sites are added to plane waves, which provides an efficient description of more localized components of the valence states. Here, plane waves with a cut-off for the kinetic energy of 24 Ry and local functions of *s, p, d* type for La and Ag, and *s, p* type for Sb, respectively, were employed. The self-consistent Kohn-Sham potential was obtained using a tetragonal 24x24x12 *k*-point mesh in conjunction with a Gaussian broadening of 50 meV. For Fermi surfaces and susceptibilities, band energies were evaluated on a much denser 100x100x24 *k*-point mesh. Interestingly here, we note that the tetragonal phase is stable against CDW instabilities in the calculation (that is all phonon frequencies remain real). As we shall see, the input from the calculation can be used to analyse the DS data and unveil the origin of the CDW formation in this system.

The DS study covered the temperature range from 100 K to room temperature (RT) with adaptive step from 5 to 20 K. The DS signal is seen in the entire reciprocal space, but can be best studied next to very weak Bragg reflections. In Fig. 1, we show representative reconstructions of the reciprocal space around the $\Gamma_{300} = (3\ 0\ 0)$ Bragg reflection in various k-space planes. In the reciprocal (h k 0) plane, at RT the DS intensity is distributed within a diagonal cross centered around $\Gamma_{300}$, and forms hollow diffuse tubes along the c* direction. These are not perfect cylinders and flatten along a*. Intensity maxima at the modulation vectors $\tau_1$ and $\tau_2$ are already clearly visible at RT.

We first focus on the formation of the CDW1 which occurs in the (h k 0) plane. Upon cooling, the DS intensity at $\tau_1$ strongly increases, resulting in a set of sharp satellite CDW reflections below $T_{CDW1}$. At 100K, these in-plane satellites alongside higher harmonics can be also clearly distinguished just above the second CDW transition along the (3 k 0) and (h 0 0) lines (Fig. 1, top panel).

On further cooling, a second set of CDW reflections - associated with CDW2 - appears at commensurate value $\tau_2 = c^*/6$. At the lowest investigated temperature (100 K) incommensurate satellites up to 6$^{th}$ order are observable and combined satellites $n\tau_1+m\tau_1'+p\tau_2$ are also clearly apparent ($\tau_1$ and $\tau_1'$ are related by $\pi/2$ rotation around $c^*$).

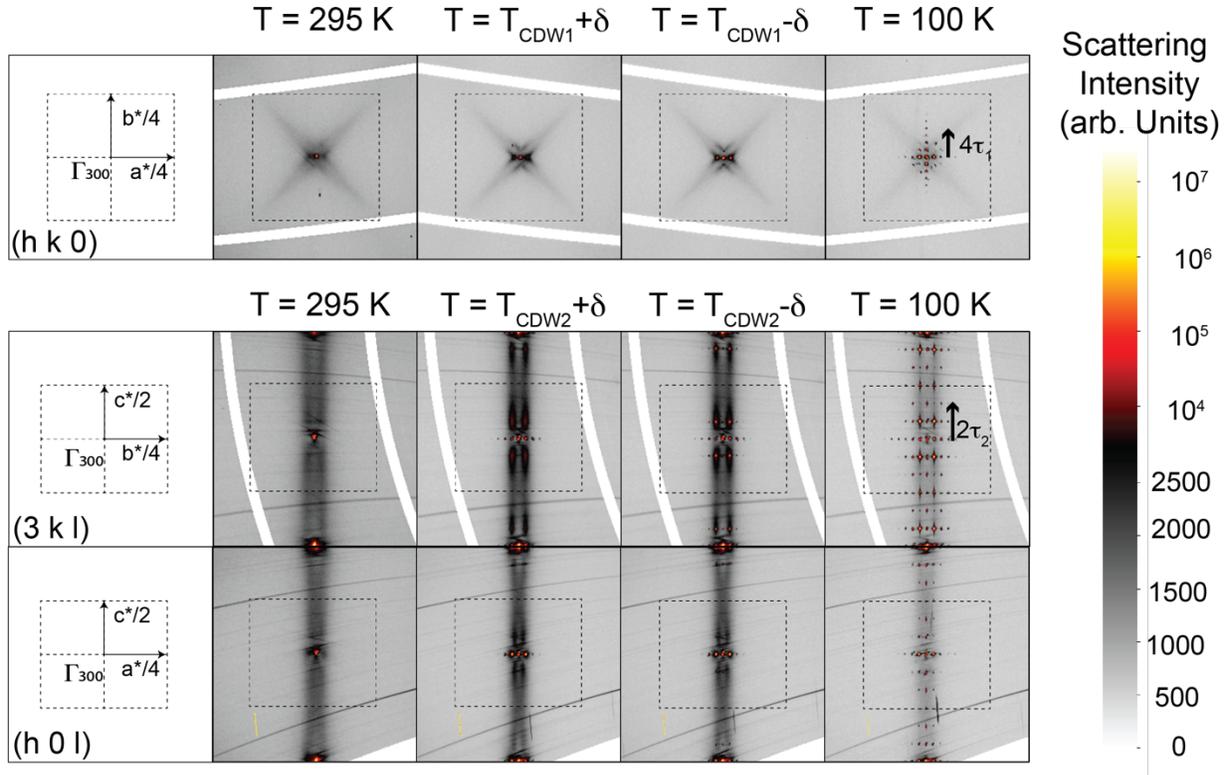

*Figure 1. Three orthogonal cuts of the LaAgSb$_2$ reciprocal space across the $\Gamma_{300}$ Bragg node as a function of temperature. $\Gamma_{300}$ is situated in the center of each panel. For clarity, multiples of the CDW modulation wave vectors $\tau_1$ and $\tau_2$ are shown. Measurement above and below the CDW transitions were taken within a temperature range of $\delta \sim 2$ K.*

In order to check whether the temperature dependence of the DS is of static (*i.e.* related to *e.g.* an order/disorder transition) or dynamic origin (phonon softening), we have carried out a series of energy-resolved IXS measurements. In order to minimize the contribution of the elastic signal arising from the central Bragg reflection, we have performed the inelastic scans across the diffuse tube wall in the b* direction, starting from $\Gamma_{300} + \tau_2 = (3\ 0\ 1/6)$. All the spectra shown in Fig. 2-a have been recorded just above T$_{CDW1}$ and are dominated by the inelastic scattering signal from a low-energy phonon. Furthermore, the mode softens upon approaching $\tau_1'$ so that Stokes and anti-Stokes peaks nearly merge for $(3\ 0\ 1/6) + \tau_1'$, where the DS intensity (Fig.1) shows a maximum. While the relatively low count rate for the chosen extreme values of energy and momentum resolution preclude systematic explorations as a function of temperature, our IXS results allow to conclude that: i) the DS is essentially of inelastic nature and ii) the maximum of intensity corresponds to the lowest frequency of the phonons.

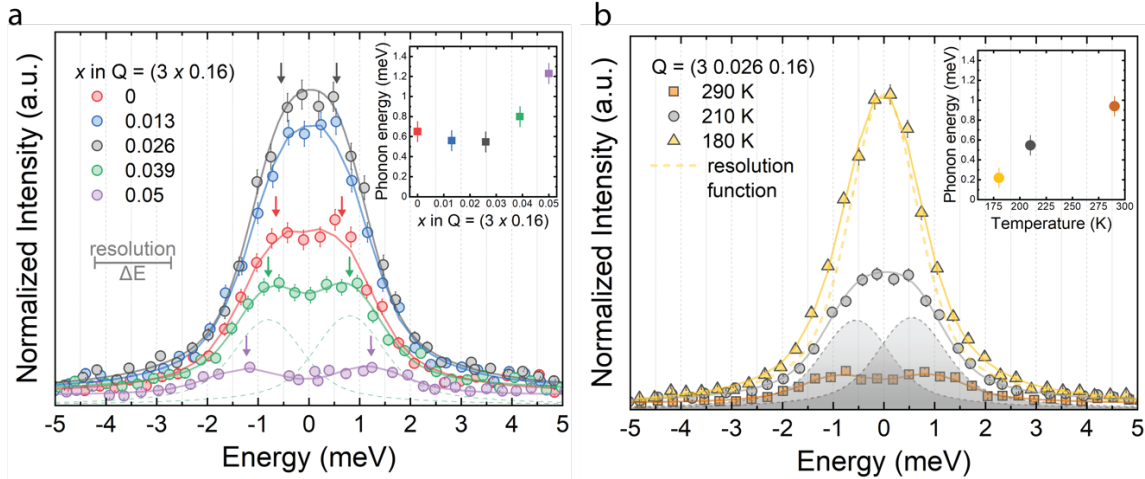

*Figure 2. a) IXS spectra taken for LaAgSb$_2$ at 210 K along (3 x 1/6) line. Insert: dispersion of the soft mode at 210K. Fit (solid lines) corresponds to the sum of damped harmonic oscillators weighted by a Bose factor (which is different for the Stokes and anti-Stokes contributions), convoluted with the experimental resolution function (for clarity, we show the decomposition as dotted lines for x=0.039 only). The arrows indicate the phonon energy as extracted from the fits. b) Temperature dependence of the IXS scan at the CDW vector (3 0.026 1/6).*

We have therefore unambiguously established that the strong temperature dependence of the DS above the CDW transition originates from the softening of phonons, which drives the CDW instabilities in LaAgSb$_2$. Without any doubt, this strong anharmonicity is rooted in the EPC, but the origin of its momentum dependence remains to be clarified. In particular, given the quasi-two dimensionality of the system it is legitimate to wonder whether those can be associated with the Dirac-like band, as suggested in previous studies [26, 29].

We argue below that a direct comparison between the momentum distribution of the DS intensity and the calculated electronic susceptibility of the normal state out-of-which the CDW emerge provides a straightforward way to settle this issue [40].

To do this, we have first calculated the real part of the static generalized electronic susceptibility:

$$\chi(\boldsymbol{q}) = \sum_{i,j=1}^{4} \chi^{ij}(\boldsymbol{q}),$$

where $\chi^{ij}(\boldsymbol{q}) = \sum_{\boldsymbol{k}} \frac{n_F[E^i(\boldsymbol{k})] - n_F[E^j(\boldsymbol{k+q})]}{E^i(\boldsymbol{k}) - E^j(\boldsymbol{k+q})}$ corresponds to a 'partial' susceptibility considering only the intra ($i = j$) and inter- ($i \neq j$) band contributions from each of the four bands crossing the Fermi level to the total susceptibility ($n_F(E)$ is the Fermi-Dirac distribution, and the energy dispersion for each band is obtained from our DFT calculation). We recall here that the imaginary part of $\chi(\boldsymbol{q})$, which involves only the electronic states at the Fermi level vanishes in the static limit due to causality [8]. Owing to the Kramers-Kronig relations, all electronic states from the dispersion contribute to the real part of $\chi(\boldsymbol{q})$ which can therefore exhibit a richer momentum structure than that infered from nesting of the states at the Fermi level only.

As can be seen in Fig. 3b, the momentum dependence of $\chi(\boldsymbol{q})$ is indeed quite complex, and in particular much richer than the DS patterns seen in Fig. 1. This is not surprising as DS does not directly probe $\chi(\boldsymbol{q})$.

On the other hand, a very different picture emerges as we inspect the individual $\chi^{ij}(\boldsymbol{q})$ contributions to $\chi(\boldsymbol{q})$. The largest intraband contributions to the $\chi(\boldsymbol{q})$ are $\chi^{11}(\boldsymbol{q}), \chi^{22}(\boldsymbol{q})$ and $\chi^{33}(\boldsymbol{q})$ and are reported in Figs. 3d, f and h. They arise from three of the bands crossing the Fermi level and the corresponding Fermi surfaces are represented in Figs. 3c, e and g, respectively. These are much more intense than $\chi^{44}(\boldsymbol{q})$ or than any of the interband contributions (See Appendix A). All these $\chi^{ij}(\boldsymbol{q})$ have pronounced and specific momentum dependences but, remarkably, only $\chi^{33}(\boldsymbol{q})$ shows a cross-like shape in the (h k 0) plane and a tubular structure along the c-axis, reminiscent of the DS patterns of Fig. 1.

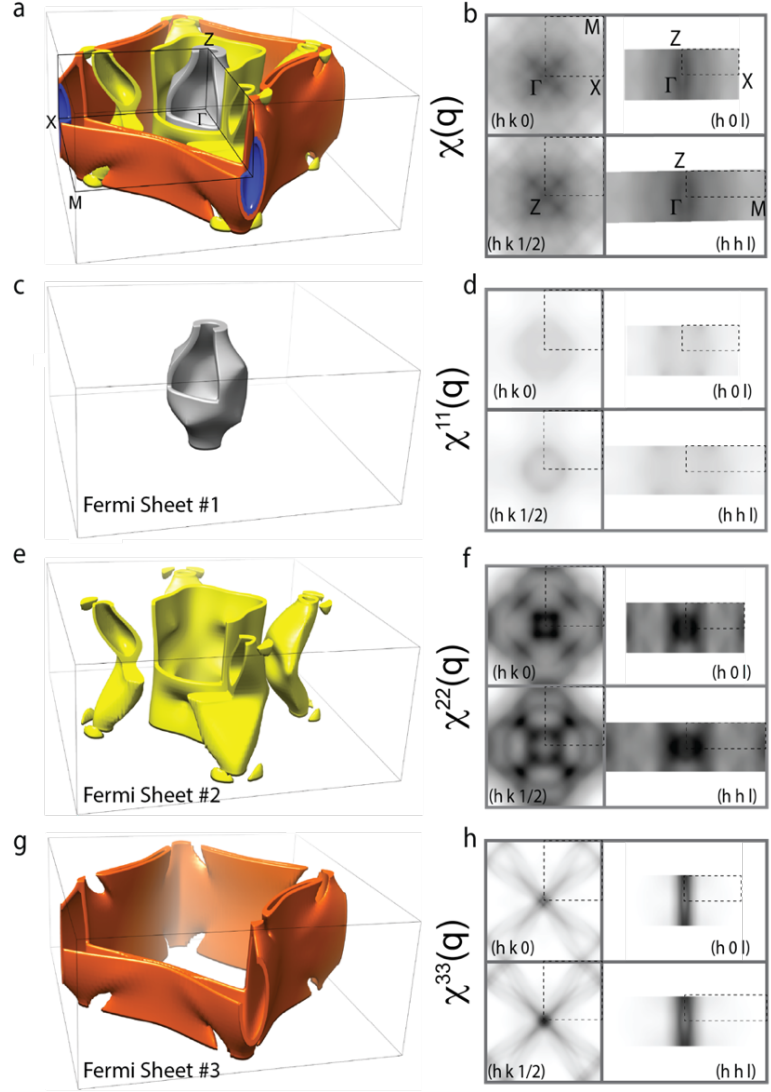

*Figure 3. Ab initio derived LaAgSb$_2$ momentum space maps (The contributions to the Fermi surface are represented as isosurfaces of $\exp\left(-E_i^2/w^2\right)$, where $E_i$ is the energy of $i^{th}$ band for given momentum and w – is a common blurring parameter): (a) four sheets of Fermi surfaces and (b) the corresponding electronic susceptibility $\chi(\mathbf{q})$ projected on various reciprocal space planes (c) Fermi surface sheet 1 and its (d) intraband electronic susceptibility $\chi^{11}(\mathbf{q})$ projected on various reciprocal space planes. (e) Fermi surface sheet 2 and its (f) intraband electronic susceptibility $\chi^{22}(\mathbf{q})$ projected on various reciprocal space planes. (g) Fermi surface sheet 3 and its (h) intraband electronic susceptibility $\chi^{33}(\mathbf{q})$ projected on various reciprocal space planes.*

*To best see the momentum structure of the $\chi^{ij}(\mathbf{q})$, we have plotted only the intensity above a constant background (corresponding to the minimal value in the Brillouin zone) using the same linear grey scale for all the $\chi^{ij}(\mathbf{q})$ (The scale for the total susceptibility is different). To see the overall corresponding intensity (without background subtraction), see panel (e) and (f) of fig. 5 where linecuts along the (1 0 0) and (1 1 0) directions of these maps are given.*

In Fig. 4, we compare in greater details the DS patterns around $\Gamma_{300}$ with the k-space structure of $\chi^{33}(\boldsymbol{q})$. The precise location of the calculated intensity maxima slightly differs from those that are observed experimentally. This reminds of the mismatch between the proposed nesting vectors and $\tau_1$ from previous studies [26, 29], but this small difference can be accounted for by a minor (~70 meV) adjustment of the Fermi level, which is reasonable compared to the absolute precision of DFT calculations (initial $E_F$ = 0.188 Ry, became 0.183 Ry). Note that the shifted Fermi surface remains in very good agreement ARPES data from ref. [29] (see Appendix).

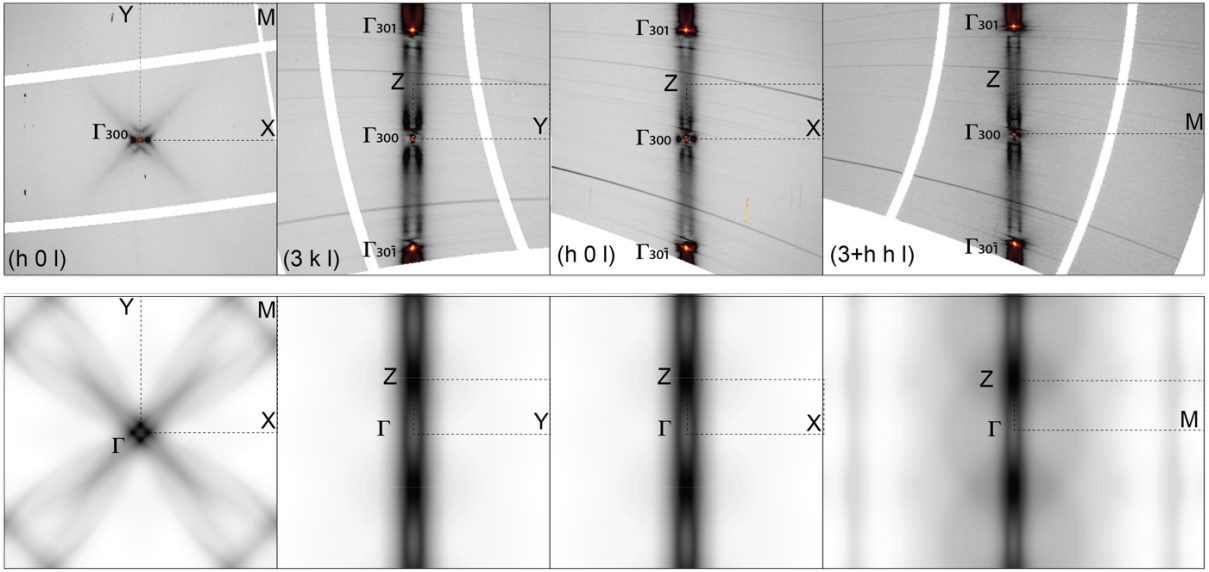

Figure 4. Comparison of DS maps of LaAgSb$_2$ at $T_{CDW1}+\delta$ ($\delta \sim 2$ K) (upper panels) with the corresponding maps of $\chi^{33}(\boldsymbol{q})$ for the band of interest (lower panels) after a 70 meV adjustment of the Fermi level.

Given the constrains imposed by these 'nesting surface shapes', which are way stronger than those derived just from the modulation vector, the excellent agreement we obtain here largely validates the adequacy of DFT calculation for the description of the band structure of LaAgSb$_2$. Importantly, all the features from our experimental data can be directly associated with $\chi^{33}(\boldsymbol{q})$ and, conversely, none of them can be found in the other $\chi^{ij}(\boldsymbol{q})$ contributions. Note finally the small mismatch of the diffuse tube diameter in two orthogonal directions (a* and b* - Fig. 4), which is not seen in $\chi^{33}(\boldsymbol{q})$ is naturally explained by the fact that the DS pattern is decorated by scattering selection rules from phonons[41], from which the fourfold symmetry is necessarily lost.

This observation immediately raises the question of why should $\chi^{33}(\boldsymbol{q})$ contribute to DS more than any other $\chi^{ij}(\boldsymbol{q})$?

We show that this naturally arises from the momentum dependence of the EPC, which we evaluate through that of the total (that is, summed over all branches) phonon linewidth $\gamma^{ij}(\boldsymbol{q}) = 2\pi \sum_{k,\lambda} \omega_{q,\lambda} \left|g^{q\lambda}_{k,i;k+q,j}\right|^2 \delta(E_{k,i}) \delta(E_{k+q,j})$. There $g^{q\lambda}_{k,i;k+q,j}$ is the matrix element corresponding to the scattering of an electron by a phonon from branch $\lambda$ between two points of the Fermi surface on bands $i$ and $j$ separated by $\boldsymbol{q}$ (see also Appendix D).

In Fig. 5a, we show that the overall contribution of EPC to linewidths $\gamma(\boldsymbol{q}) = \sum_{i,j} \gamma^{ij}(\boldsymbol{q})$ along the (1 0 0) reciprocal direction, exhibits a clear maximum close to $\tau_1$, which arises from $\gamma^{33}(\boldsymbol{q})$, that is from the total contribution to EPC of the electronic states that form the Fermi surface represented in Fig. 3g. It is interesting to note that the momentum structure of $\gamma^{ij}(\boldsymbol{q})$ closely follows that of the Fermi surface nesting function given by the imaginary part of the electronic susceptibility $N(\boldsymbol{q}) = \sum_{ij} N_{ij}(\boldsymbol{q})$, with $N_{ij}(\boldsymbol{q}) = \lim_{\omega \to 0} \frac{\chi^{ij''}(q,\omega)}{\omega} = \sum_k \delta(E_{k,i}) \delta(E_{k+q,j})$ as can be seen in Fig. 5c. In other words, the combination of a large electronic static susceptibility $\chi^{33}(\boldsymbol{q})$ (as seen in the line cuts on Figs. 5e and f) and of a momentum dependent EPC, both peaked near $\tau_1$ yields strong electron-phonon scattering at $\tau_1$, which results in the observed phonon softening (Fig. 2), explains the contribution of $\chi^{33}(\boldsymbol{q})$ to DS close to $\Gamma_{300}$[41] and finally drives the formation of the CDW.

On the contrary, even though other $\chi^{ij}(\boldsymbol{q})$ exhibit a strong momentum dependence (*e.g.* $\chi^{11}(\boldsymbol{q})$ or $\chi^{22}(\boldsymbol{q})$), if not combined with a matching EPC $\gamma^{ij}(\boldsymbol{q})$ that would enable the scattering of the electrons by the phonons at the relevant momenta (and thereby a softening of the phonons), they will not contribute to DS. This conclusion is supported by the observation of a featureless $\gamma^{ij}(\boldsymbol{q})$ along the (1 1 0) direction (Fig. 5b), that could otherwise have triggered a CDW instability in this direction, in which $\chi^{22}(\boldsymbol{q})$ is for instance particularly strong.

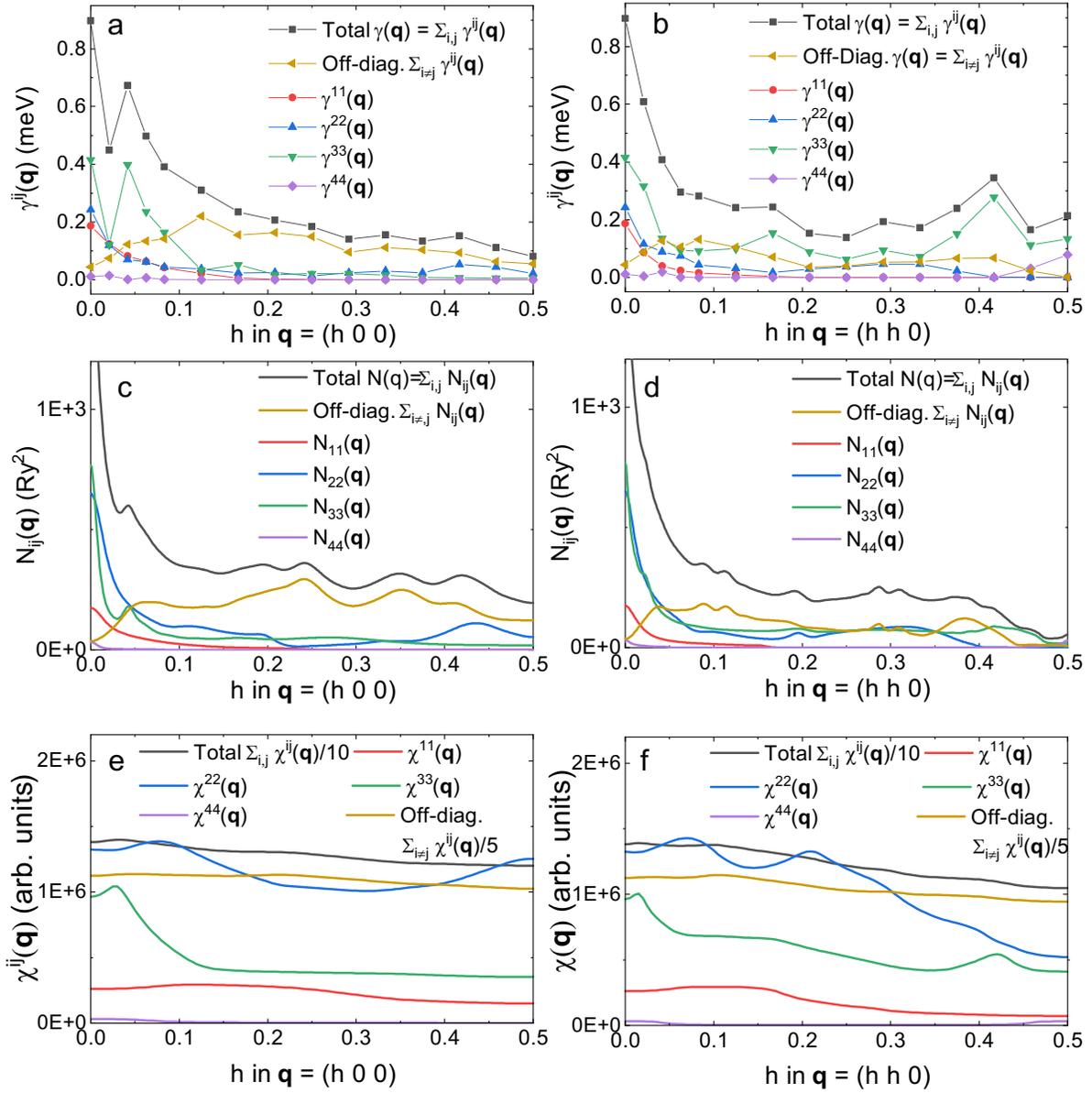

*Figure 5 (a) Total phonon linewidth along the (1 0 0) direction obtained by summing the contribution of EPC to all phonon branches (see text). (b) Total phonon linewidth along the (1 1 0) direction obtained by summing the contribution of EPC to all phonon branches (see text). (c) Fermi Surface nesting functions $N_{ij}(q)$ along the (1 0 0) direction. (d) Fermi Surface nesting function along the (1 1 0) direction. In each panel, the contributions from the different bands are isolated (the interband, off diagonal, contributions are summed). (e) Line cuts of the susceptibilities $\chi^{ij}(q)$ along the (1 0 0) direction. (f) Line cuts of the susceptibilities $\chi^{ij}(q)$ along the 110 direction. In panels (e) and (f) the total susceptibility has been divided by 10 and the off-diagonal contribution by 5.*

Taken together, these facts allow us to definitively claim that the formation of the CDW in LaAgSb$_2$ is directly related to a nesting mechanism in which anisotropic EPC is rooted. It is important to highlight that this conclusion could not have been drawn from the analysis of the total electronic susceptibility $\chi(\boldsymbol{q})$, in which the specific contribution of $\chi^{33}(\boldsymbol{q})$ is averaged out. In other words, in a multiband system such as LaAgSb$_2$, it is absolutely crucial to examine each $\chi^{ij}(\boldsymbol{q})$ contribution individually, as well as the momentum dependence of the corresponding EPC $\gamma^{ij}(\boldsymbol{q})$ - rather than the total $\chi(\boldsymbol{q})$ - to assess the relevance of nesting for the formation of the CDW.

We finally note that a comparison between Figs 1 and 4 also reveals that the tubular shape of the DS pattern along the c-axis is not affected by the soft-phonon condensation at $T_{CDW1}$. This tubular structure is only visible in $\chi^{33}(\boldsymbol{q})$, which is therefore also responsible for the formation of the CDW2 (but a different phonon branch is likely involved – its identification is however beyond the scope of the present study).

We end our discussion by noting that our result might appear in contradiction with the common wisdom that nesting is not relevant in d>1 materials [8, 9]. Detailed analysis of the dispersion of the electronic states shown in Fig. 3g, which originate from the 5p$_x$ and 5p$_y$ states of Sb, reveals a linear dispersion from (0.192 0.192 0.5) along the (h h 0) reciprocal space direction, previously identified as Dirac feature [29]. Interestingly the electronic dispersion is steep and linear, but only in this direction and it appears flatter and parabolic in the orthogonal ones (See Appendix C). It is this combination of a steep linear dispersion near E$_F$ with a strong anisotropy that enables the strong nesting which contributes to the CDW instabilities. This provides interesting perspectives for the rational design of CDW states through that of Dirac-like features in band structure engineering.

In conclusion, we have performed DS and IXS experiments across the two CDW transitions in LaAgSb$_2$. The high quality of our crystals allowed for the observation of a very rich diffraction pattern including CDW satellites up to the 6$^{th}$ order. We demonstrated that the formation of both observed families of CDWs in LaAgSb$_2$ is driven by soft phonons. The corresponding Kohn anomaly, visualized in 3D through the diffuse scattering, is confronted to the electronic susceptibility and EPC within and between the individual bands contributing to the complex Fermi surface. The remarkable agreement between experimentally measured and calculated patterns demonstrates that the combination of Fermi surface nesting of a highly anisotropic dispersing band, linear in one direction of the reciprocal space, with a strongly momentum dependent electron-phonon interaction is directly responsible for the CDW instability in this quasi-2D material.

**Acknowledgments**

This work has been partially supported by the CNRS via French-Polish International Research Project IRP '2D Materials' and by the EC Graphene Flagship Project. NLW acknowledges the support by National Natural Science Foundation of China (No. 11888101), the National Key Research and Development Program of China (No. 2017YFA0302904, 2016YFA0300902). RH acknowledges support by the state of Baden-Württemberg through bwHPC. We thank F. Weber for commenting on the manuscript.

# APPENDIX A: intra- and interband contributions to $\chi(\boldsymbol{q})$

In Fig. 6, we show all the intra- and interband $\chi^{ij}(\boldsymbol{q}) = \sum_{\boldsymbol{k}} \frac{n_F[E^i(\boldsymbol{k})] - n_F[E^j(\boldsymbol{k}+\boldsymbol{q})]}{E^i(\boldsymbol{k}) - E^j(\boldsymbol{k}+\boldsymbol{q})}$ contributions to the total static electronic susceptibility $\chi(\boldsymbol{q}) = \sum_{i,j=1}^{4} \chi^{ij}(\boldsymbol{q})$ calculated for the four bands of LaAgSb$_2$ crossing the Fermi level and plotted in selected high-symmetry directions of the reciprocal space. As mentioned in the main text, $\chi(\boldsymbol{q})$ is largely dominated by the intraband contributions $\chi^{22}(\boldsymbol{q})$ and $\chi^{33}(\boldsymbol{q})$.

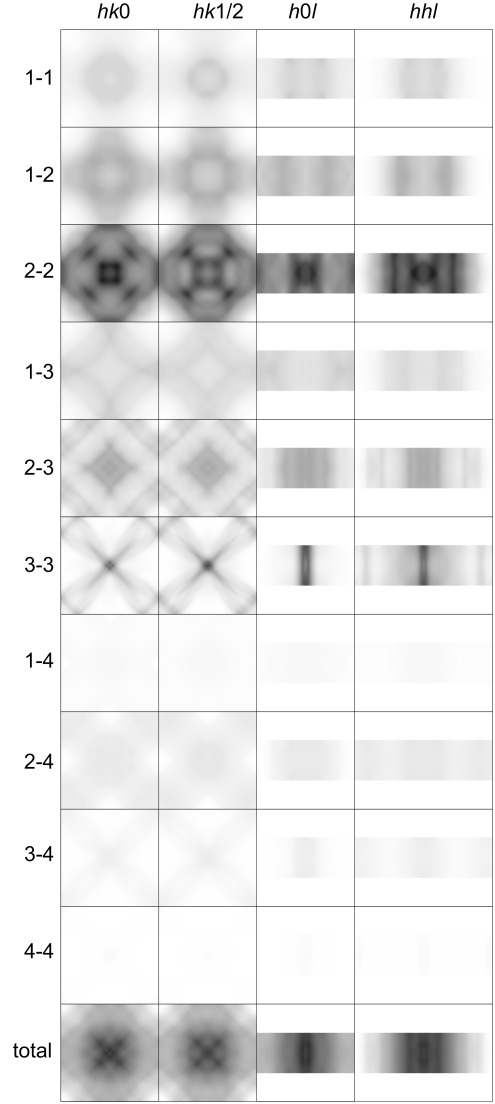

***Figure 6*** *intra- and interband contributions to the electronic susceptibility $\chi(\boldsymbol{q})$ projected on various reciprocal space planes. To best see the momentum structure of the $\chi^{ij}(\boldsymbol{q})$, we have plotted only the intensity above a constant background (corresponding to the minimal value in the Brillouin zone) using the same linear grey scale for all the $\chi^{ij}(\boldsymbol{q})$ (The scale for the total susceptibility is different).*

**APPENDIX B: Calculated Fermi surface and comparison with ARPES data**

In Fig. 7, we provide a direct comparison of our calculated Fermi surface with the experimentally determined one from ref. [29]. We show two cuts of the Fermi surface in the $k_z = 0$ and $k_z = 0.5$ planes. Our calculation for the later is strikingly similar that reported in [29], suggesting that the authors of this publication used a finite $k_z$ for comparison with the experiment (for which $k_z$ is always hard to define). The agreement for the Fermi surface sheets close to the X point appears better for the $k_z = 0$ calculation, whereas the central pocket shape in the experiment seems closer to that calculated for $k_z = 0.5$. In any event, the agreement between the calculation and the experiment is very good, and is not affected by the small 70 meV Fermi level shift we applied to reproduce our diffuse scattering data.

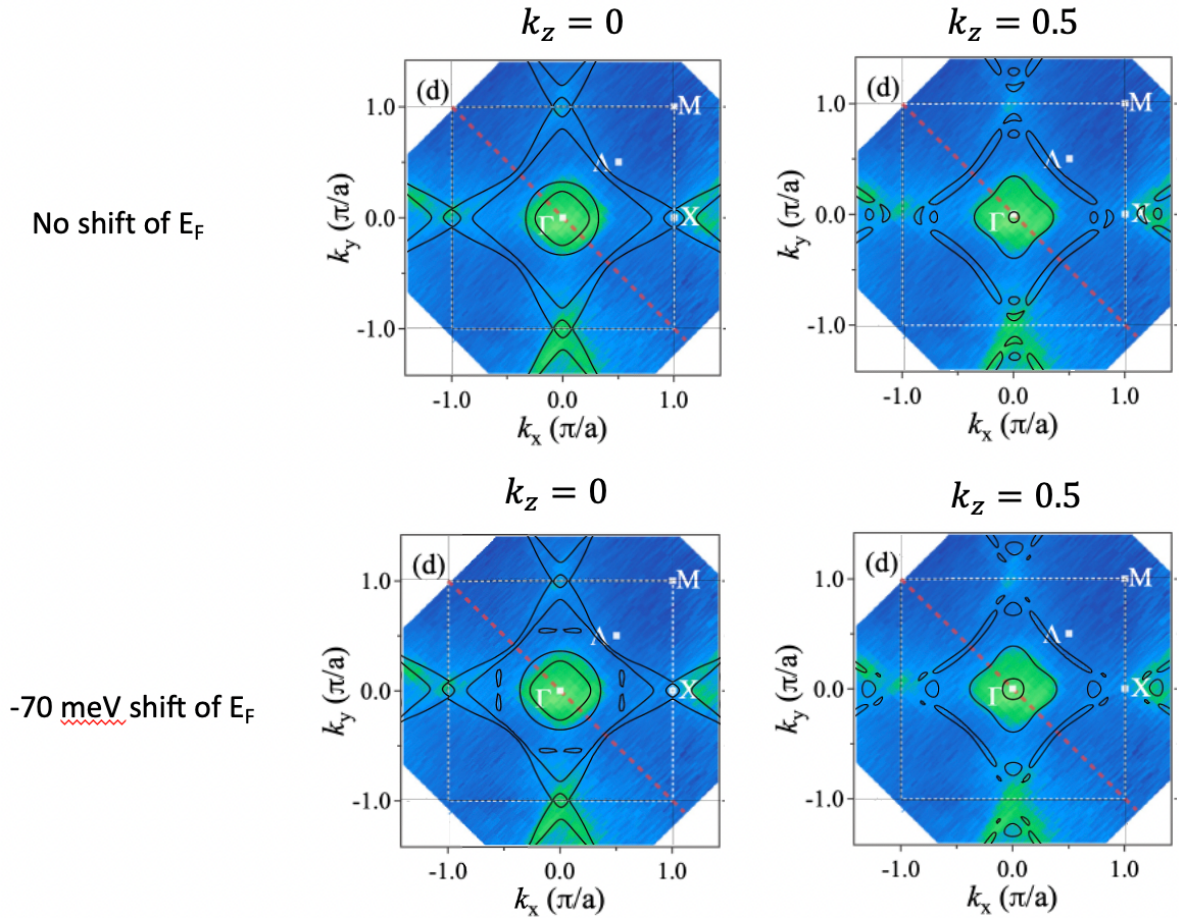

*Figure 7 Comparison between our calculated Fermi surface (black lines) with that experimentally measured in [29]. We show two cuts of the Fermi surface in the $k_z= 0$ and $k_z=0.5$ planes for the original calculation (top line), and that for $E_F$ shifted by 70meV (bottom line), from which we obtain the best agreement with our diffuse scattering data.*

## APPENDIX C: Dirac-like point in LaAgSb$_2$

In Fig. 8, we show the electronic band structure of LaAgSb$_2$ in the vicinity of the Dirac-like points. Band crossings with linear dispersive bands appear along the high-symmetry lines M-Γ and A-Z at points K$_1$=(k$_1$,k$_1$,0) with k$_1$=0.1920 and K$_2$=(k$_2$,k$_2$,0.5) with k$_2$=0.2028, respectively (see panels (a) and (b)). The linear dispersion is, however, not observed along directions orthogonal to these high-symmetry lines. This can be seen in panels (c) and (d), where we display the band dispersion along lines through the crossing points parallel to the z-axis, as well as in panels (e) and (f) along an orthogonal direction within the xy-plane (δ=0.05). Along these orthogonal lines, all bands exhibit a quadratic dispersion. This demonstrates that the band structure of LaAgSb$_2$ does not possess a true 3D Dirac cone with linear dispersing bands in all directions.

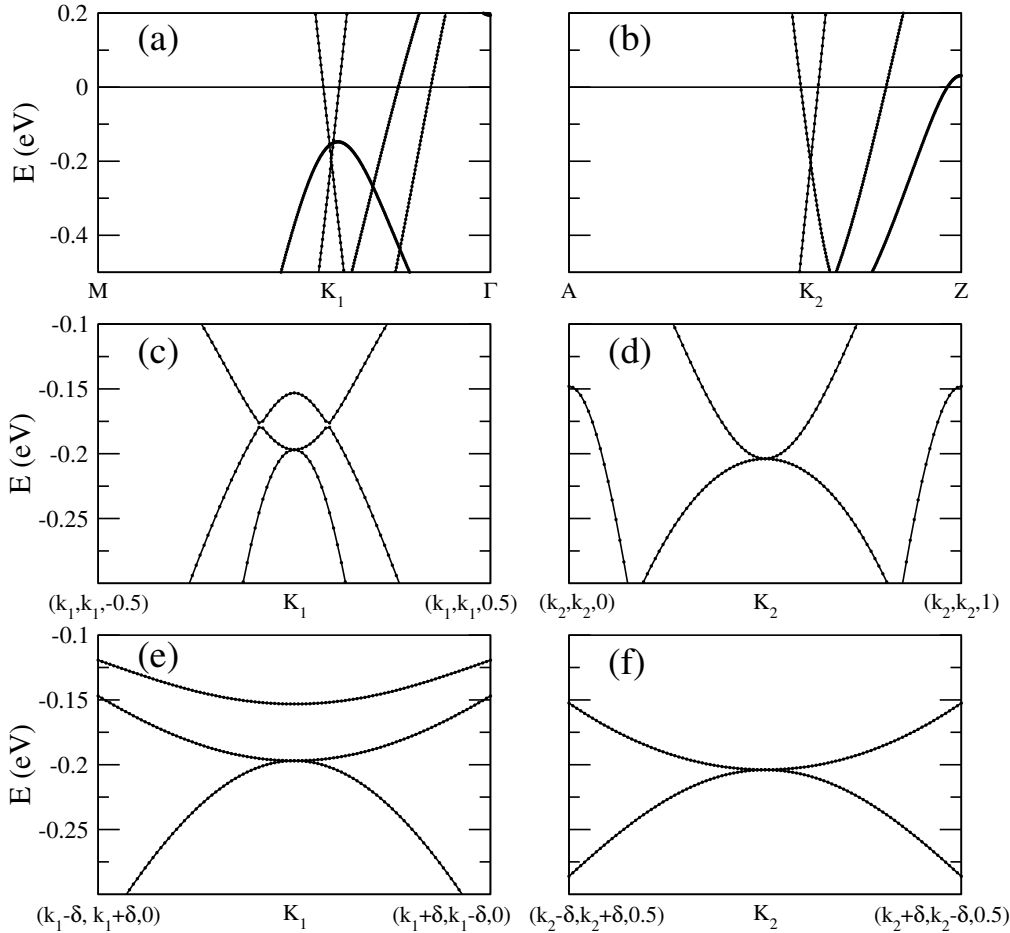

*Figure 8* electronic band structure of LaAgSb$_2$ in the vicinity of the Dirac-type points. Band crossings with linear dispersive bands appear along the high-symmetry lines M-Γ and A-Z at points a) K$_1$=(k$_1$,k$_1$,0) with k$_1$=0.1920 and b) K$_2$=(k$_2$,k$_2$,0.5) with k$_2$=0.2028, respectively. (c) band dispersion along parallel to the z-axis across K1 (c) and K2 (d), and along an orthogonal direction in the xy-plane (δ=0.05).

**APPENDIX D: Phonon dispersion and structure factors**

Phonon frequencies and eigenvectors were obtained via density functional perturbation theory as implemented in the mixed-basis scheme [42]. In addition, this approach provides direct access to the scattering potential induced by a phonon, which is then subsequently used to calculate EPC matrix elements. To achieve convergence, phonon linewidths $\gamma^{ij}(\boldsymbol{q}) = 2\pi \sum_{\boldsymbol{k},\lambda} \omega_{\boldsymbol{q},\lambda} \left| g^{q\lambda}_{\boldsymbol{k},i;\boldsymbol{k}+\boldsymbol{q},j} \right|^2 \delta(E_{\boldsymbol{k},i}) \delta(E_{\boldsymbol{k}+\boldsymbol{q},j})$ were evaluated by performing the k sum over a dense 48x48x24 tetragonal mesh, while replacing the delta functions by Gaussians with a width of 50 meV. In the previous expression, $g^{q\lambda}_{\boldsymbol{k},i;\boldsymbol{k}+\boldsymbol{q},j}$ stands for the EPC matrix element corresponding to the scattering of an electron by a phonon from branch $\lambda$ between two points of the Fermi surface on bands $i$ and $j$ separated by $\boldsymbol{q}$. It is evaluated as the matrix element of the (screened) first-order change of the crystal potential, as induced by a phonon mode, and two electronic (Kohn-Sham) states $g^{q\lambda}_{\boldsymbol{k},i;\boldsymbol{k}+\boldsymbol{q},j} = \langle \boldsymbol{k}+\boldsymbol{q},j | \delta_{q\lambda} V | \boldsymbol{k},i \rangle$. The potential change $\delta_{q\lambda} V$ is directly obtained in density functional perturbation theory. The resulting phonon dispersion along the (1 0 0) direction is plotted in the Fig. 9, in which the vertical error bars represent the phonon linewidths. The color map represents the scattering intensity in the Brillouin zone centered around $\Gamma_{300}$, where our IXS measurements have been carried out in transverse geometry. A clear Kohn anomaly of several optical branches is seen but the modes are stables.

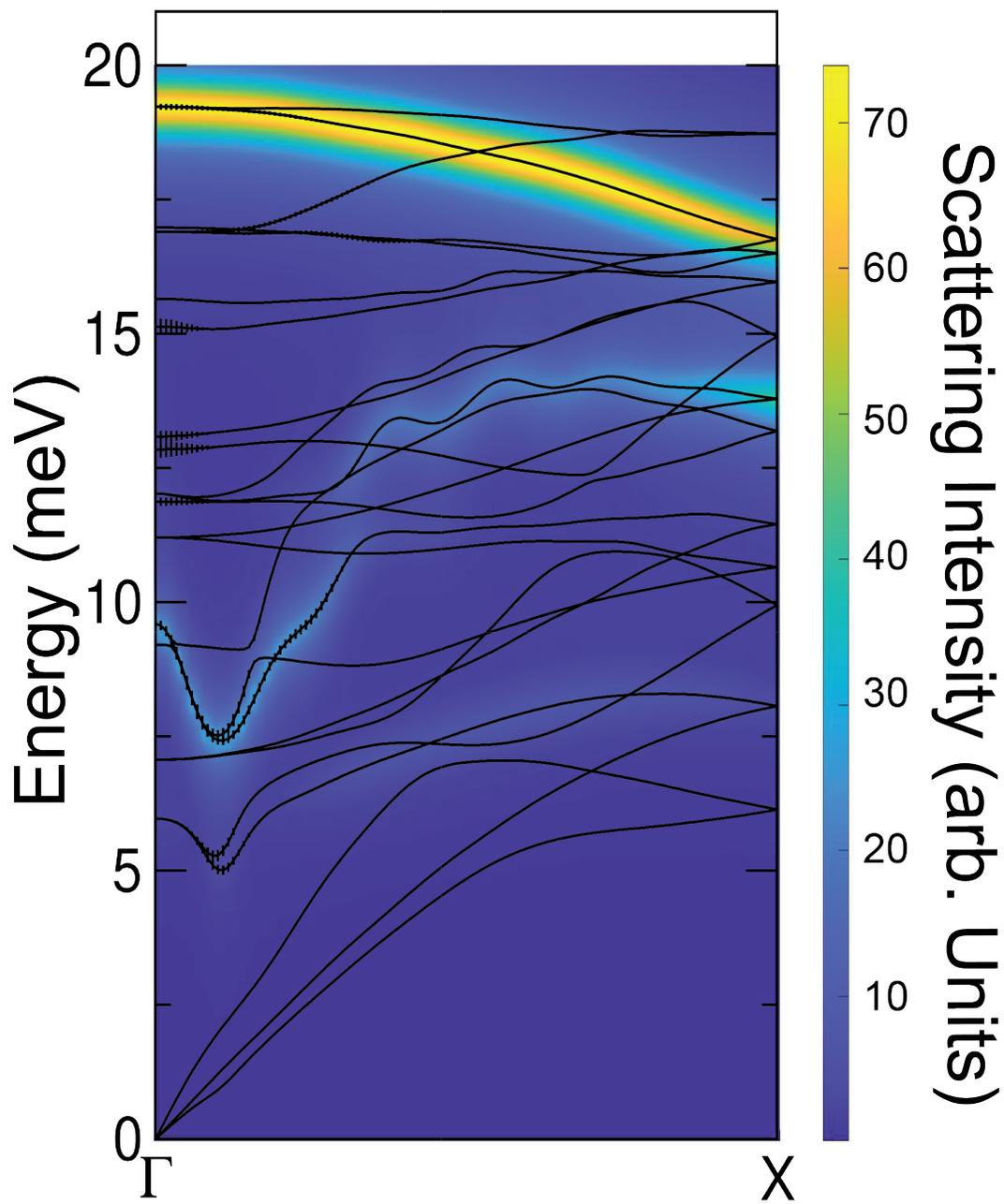

*Figure 9* Calculated phonon dispersion for LaAgSb$_2$ along the high-symmetry line ΓX. The vertical error bars represent the phonon linewidth, and the color plot the scattering intensity calculated along the (3 k 0) line.

## References

[1] J.P. Pouget, S. Kagoshima, C. Schlenker, and J. Marcus, Journal de Physique Lettres, Edp sciences, **44**, 113 (1983).
[2] F. Denoyer, R. Comès, A. F. Garito, and A. J. Heeger, Phys. Rev. Lett. **35**, 755 (1975).
[3] M. Hoesch, A. Bosak, D. Chernyshov, H. Berger, and M. Krisch, Phys. Rev. Lett. **102**, 086402 (2009).
[4] L. Yue, S. Xue, J. Li, W. Hu, A. Barbour, F. Zheng, L. Wang, J. Feng, S. B. Wilkins, C. Mazzoli, R. Comin, and Y. Li, Nature Communications **11**, 98 (2020).
[5] J.A. Wilson, F.J. Di Salvo, and S. Mahajan Adv. Phys. **24**, 117 (1975).
[6] G. Ghiringhelli, M. Le Tacon, M. Minola, S. Blanco-Canosa, C. Mazzoli, N.B. Brookes, G.M. De Luca, A. Frano, D. G. Hawthorn, F. He, T. Loew, M. Moretti Sala, D.C. Peets, M. Salluzzo, E. Schierle, R. Sutarto, G. A. Sawatzky, E. Weschke, B. Keimer, and L. Braicovich, Science **337**, 821 (2012).
[7] M. Le Tacon, A. Bosak, S. M. Souliou, G. Dellea, T. Loew, R. Heid, K. P. Bohnen, G. Ghiringhelli, M. Krisch, and B. Keimer, Nature Physics **10**, 52 (2014).
[8] S. Lee, G. de la Peña, S. X.-L. Sun, M. Mitrano, Y. Fang, H. Jang, J.-S. Lee, C. Eckberg, D. Campbell, J. Collini, J. Paglione, F. M. F. de Groot, and P. Abbamonte, Phys. Rev. Lett. **122**, 147601, (2019).
[9] R. Peierls, Quantum Theory of Solids, Oxford Univ Press, NY (1955)
[10] M. D. Johannes and I. I. Mazin, Phys. Rev. B **77**, 165135 (2008).
[11] X. Zhu, Zhu, Y. Cao, J. Zhang, E. W. Plummer, and J. Guo, Proceedings of the National Academy of Sciences **112**, 2367 (2015).
[12] K. Rossnagel, Journal of Physics: Condensed Matter **23**, 213001 (2011).
[13] V. N. Strocov, M. Shi, M. Kobayashi, C. Monney, X. Wang, J. Krempasky, T. Schmitt, L. Patthey, H. Berger, and P. Blaha, Phys. Rev. Lett. **109**, 086401 (2012).
[14] F. Weber, S. Rosenkranz, J. P. Castellan, R. Osborn, R. Hott, R. Heid, K. P. Bohnen, T. Egami, A. H. Said, and D. Reznik, Phys. Rev. Lett. **107**, 107403 (2011).
[15] F. Weber, S. Rosenkranz, J. P. Castellan, R., Osborn, G., Karapetrov, R., Hott, R., Heid, K. P. Bohnen, and A. Alatas, Phys. Rev. Lett. **107**, 266401 (2011).
[16] H.-M. Eiter, M. Lavagnini, R. Hackl, E. A. Nowadnick, A. F. Kemper, T. P. Devereaux, J.-H. Chu, J. G. Analytis, I. R. Fisher, and L. Degiorgi Proceedings of the National Academy of Sciences **110** 64 (2013).
[17] Felix Flicker and Jasper van Wezel, Nature Communications **6**, 7034 (2015).
[18] M. Leroux, I. Errea, M. Le Tacon, S. M. Souliou, G. Garbarino, L. Cario, A. Bosak, F. Mauri, M. Calandra, and P. Rodière, Phys. Rev. B **92**, 140303 (2015).
[19] M. Leroux, M. Le Tacon, M. Calandra, L. Cario, M. A. Measson, P. Diener, E. Borrissenko, A. Bosak, and P. Rodière, Phys. Rev. B **86**, 155125 (2012).
[20] C. Heil, S. Poncé, H. Lambert, M. Schlipf, E. R. Margine, and F. Giustino, Phys. Rev. Lett. **119**, 087003 (2017).
[21] J. Diego, A. H. Said, S. K. Mahatha, R. Bianco, L. Monacelli, M. Calandra, F. Mauri, K.Rossnagel, I. Errea, and S. Blanco-Canosa ArXiv:2007.08413 (2020).
[22] K. B. Efetov, H. Meier, and C. Pepin, Nat. Phys. **9**, 442 (2013).
[23] A. Kogar, S. Rak, S. Vig, A. A. Husain, F. Flicker, Y. Il Joe, L. Venema, G. J. MacDougall, T. C. Chiang, E. Fradkin, J. van Wezel, and P. Abbamonte, Science **358**, 1314 (2017).

[24] K.D. Myers, S.L. Bud'ko, I.R. Fisher, Z. Islam, H. Kleinke, A.H. Lacerda, and P. C. Canfield, J. Magn. Magn. Mater. 205, 27 (1999).
[25] K. D. Myers, S. L. Bud'ko, V. P. Antropov, B. N. Harmon, P. C. Canfield, and A. H. Lacerda Phys. Rev. B **60**, 13371 (1999).
[26] C. Song, J. Park, J. Koo, K.-B. Lee, J.Y. Rhee, S.L. Bud'ko, P.C. Canfield, B.N. Harmon, and A. I. Goldman, Phys. Rev. B **68**, 035113 (2003)
[27] C.S. Lue, Y.F. Tao, K.M. Sivakumar, and Y.K. Kuo, Journal of Physics: Condensed Matter **19**, 406230 (2007).
[28] K. Wang and C. Petrovic, Phys. Rev B **86**, 155213, (2012).
[29] X. Shi, P. Richard, K. Wang, M. Liu, C.E. Matt, N. Xu, R.S. Dhaka, Z. Ristic, T. Qian, Y.-F. Yang, C. Petrovic, M. Shi, and H. Ding, Phys. Rev. B **93**, 081105 (2016).
[30] R. Y. Chen, S. J. Zhang, M. Y. Zhang, T. Dong, and N. L. Wang, Phys. Rev. Lett. **118**, 107402 (2017).
[31] A. Girard, T. Nguyen-Thanh, S.-M. Souliou, M. Stekiel, W. Morgenroth, L. Paolasini, A. Minelli, D. Gambetti, B. Winkler, A. Bosak, Journal of Synchrotron Radiation **26**, 272 (2019).
[32] https://www.rigaku.com/en/rigakuoxford
[33] M. Krisch and F. Sette, Inelastic X-ray Scattering from Phonons, in Light Scattering in solids, Novel Materials and Techniques, volume 108 of Topics in Applied Physics, Springer-Verlag, 2007.
[34] https://www.oxcryo.com/
[35] S. G. Louie, K.-M. Ho, and M. L. Cohen, Phys. Rev. B **19**, 1774 (1979).
[36] B. Meyer, C. Elsässer, M. Fähnle, FORTRAN90 Program for Mixed-Basis Pseudopotential Calculations for Crystals, Max-Planck-Institut für Metallforschung, Stuttgart (unpublished).
[37] J. P. Perdew and Y.Wang, Phys. Rev. B **45**, 13244 (1992).
[38] D. Vanderbilt, Phys. Rev. B **32**, 8412 (1985).
[39] M. Brylak, M. H. Möller, and W. Jeitschko, J. Solid State Comm. **115**, 305 (1995).
[40] A. Bosak, M. Hoesch, M. Krisch, D. Chernyshov, P. Pattison, C. Schulze-Briese, B. Winkler, V. Milman, K. Refson, D. Antonangeli, and D. Farber, Phys. Rev. Lett. **103**, 076403 (2009).
[41] A. Bosak, D. Chernyshov, B. Wehinger, B. Winkler, M. Le Tacon, and M. Krisch, Journal of Physics D: Applied Physics **48**, 504003 (2015).
[42] R. Heid, K.-P. Bohnen, Phys. Rev. B 60, R3709 (1999).